\documentclass[journal,a4paper]{IEEEtran}

\usepackage{enumerate}

\newtheorem{defn}{Definition}
\usepackage{cite}

\ifCLASSINFOpdf
\else
   \usepackage[dvips]{graphicx}
   \DeclareGraphicsExtensions{.eps}
\fi
\usepackage[cmex10]{amsmath}
\usepackage{amsfonts}
\usepackage{amssymb}
\usepackage{supertabular}

\usepackage[tight,footnotesize]{subfigure}
\usepackage[colorlinks=true, citecolor=blue,urlcolor=blue, linkbordercolor={0 0 1}]{hyperref}
\usepackage{breakurl}

\begin{document}
%
\title{Yet Another Pseudorandom Number Generator}
%
%

\author{Borislav~Stoyanov, Krzysztof~Szczypiorski, and~Krasimir~Kordov
\thanks{Borislav Stoyanov and Krasimir Kordov are with the Department of Computer Informatics, Konstantin Preslavsky University of Shumen, 9712 Shumen, Bulgaria, e-mails: borislav.stoyanov@shu.bg, krasimir.kordov@shu.bg.}
\thanks{Krzysztof~Szczypiorski is with Warsaw University of Technology, Warsaw,  Poland;  Cryptomage  SA,  Wroclaw,  Poland  (e-mail: ksz@tele.pw.edu.pl)}
}

%
%

%
\markboth{}{}
%
%



%
\pagestyle{empty}%
\maketitle%
\thispagestyle{empty}%

\begin{abstract}
We propose a novel pseudorandom number generator based on R\"ossler attractor and bent Boolean function. We estimated the output bits properties by number of statistical tests. The results of the cryptanalysis show that the new pseudorandom number generation scheme provides a high level of data security.
\end{abstract}

\begin{IEEEkeywords}
R\"ossler attractor, bent Boolean function, pseudorandom number generator.
\end{IEEEkeywords}

%
\IEEEpeerreviewmaketitle

\section{Introduction}

%
%
%
%
\IEEEPARstart{B}{oolean} and chaotic functions have been used extensively in the area of a pseudorandom number generations. 

Novel encryption scheme based on bent Boolean function and feedback with carry
shift register is proposed in \cite{StoyanoKordovBentFCSR}.  In \cite{StoyanovLorenzBent}, a cryptographic algorithm based on the Lorenz chaotic attractor and 32 bit bent Boolean function is presented.

In \cite{Dascalescu}, a new chaotic system with good cryptographic properties, is proposed. Novel pseudorandom generation algorithm based on Chebyshev polynomial and Tinkerbell map, is provided in \cite{StoyanovChebTink}. Pseudorandom bit generators, based on the Chebyshev map and rotation equation, are proposed in \cite{StoyanovCheb}, \cite{StoyanovKordovEntropy2015}, and \cite{StoyanovKordovChebDuff}. The presented schemes exhibit high level of security. The proposed scheme shows that the output stream possesses suitable properties for security-demanding applications. A modified pseudorandom bit generator, based on Tinkerbell map, is presented in \cite{MalchevIbryam2015}. Pseudorandom zero-one generation algorithm based on two chaotic Circle maps and XOR function is designed in \cite{KordovCircleXOR}. In \cite{KordovSignature}, pseudorandom number generation scheme, based on Signature attractor is presented. 

Pseudorandom bit generation algorithms, based on Circle map and based on Chirikov map are proposed in \cite{StoyanovCircle} and in \cite{StoyanovKordovChirJabri}.

Several scientific papers present cryptographic primitives build from R\"ossler attractor. 

An algorithm for secure data transmission with R\"ossler function protection of the
information signal is presented in \cite{Chantov}. In \cite{FrunzeteLNCS},  an improvement to an existing algorithm used in security data sending by modifying the structure of the R\"ossler map behaviour, is designed.

A fingerprint image encryption method based on hyperchaotic R\"ossler map is provided in \cite{abundiz2016fingerprint}. The statistical analysis is presented to prove the secrecy of the biometric trait by using novel scheme.

A communication scheme to encrypted audio and image information transmission, based on hyperchaotic generalized H{\'e}non and R\"ossler maps is designed in \cite{aguilar2008synchronization}. The scheme is suitable in master-slave configuration. Another communication scheme with high stability in the recovered signal, based on R\"ossler circuit, is presented in \cite{Garcia}.

The stability of impulsive synchronization of chaotic and R\"ossler hyperchaotic systems by using Lyapunov exponent of the variational synchronization error systems are studied in \cite{Itoh}.

An algorithm to image encryption by using a R\"ossler chaotic function is presented in \cite{al2012digital}. The  approach consists  of two  substitution methods  and  two  scrambling  methods;  to change the value of the pixels and the location of the pixels, respectively.

A chaotic permutation for physical layer security in OFDM-PON is presented in \cite{liu2014physical}. An encryption algorithm based on improved hyperchaotic R{\"o}ssler map is proposed in \cite{orozco2015image}.  The proposed scheme is  attractive for applications in private communication systems.

In \cite{Sambas2013Rossler}, design and simulation of synchronization between two identical coupled R{\"o}ossler circuits, are proposed.

In \cite{BanerjeeD}, based on the R\"ossler attractor, random sequences generator is proposed. The algorithm is SIMULINK modelled and is tested using statistical tests. Random number generator from R\"ossler attractor also is presented  in \cite{Canals}. The output chaotic signal proposes a negligible value of an autocorrelation function.

A pseudorandom number generator is proposed in this paper. The novel algorithm is based on chaotic function and bent Boolean function. The novelty of our approach lies in the combination of the R\"ossler attractor and Maiorana function.

In Section 2, we propose novel pseudorandom number generator and its security analysis. Finally, the last section concludes the article.

\section{Pseudo-random bit generator from R\"ossler attractor}
\subsection{R\"ossler chaotic attractor}
\begin{figure}[ht]
\centering
\includegraphics[width={3.5in}]{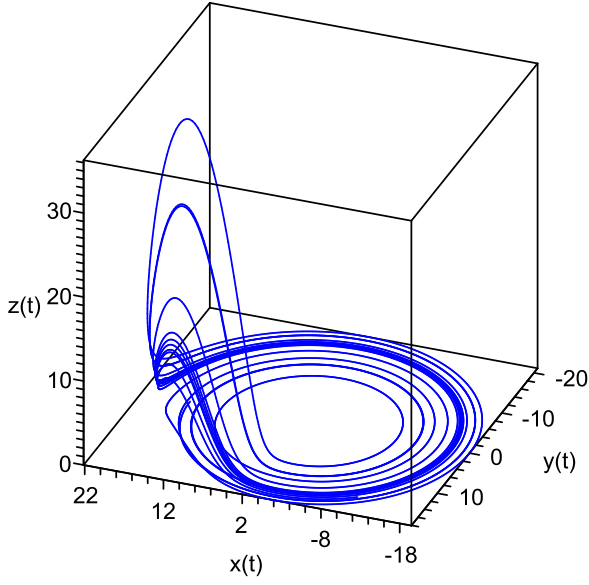}
\caption{3-Dimensional plot of R{\"o}ssler Attractor}
\label{fig:ra3d}
\end{figure}

\begin{figure}[ht]
\begin{center}
\subfigure[]{
	\includegraphics[height=.25\textheight]{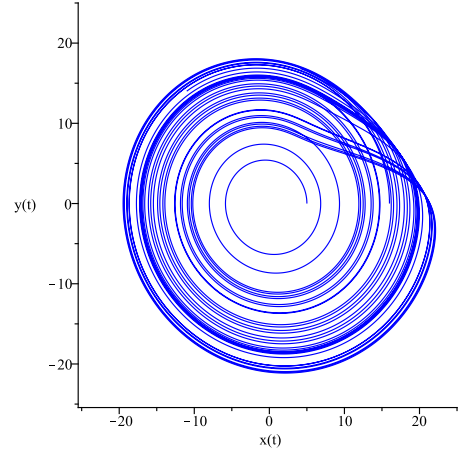}
	\label{fig:raxy}	
}
\\
\subfigure[]{
	\includegraphics[height=.24\textheight]{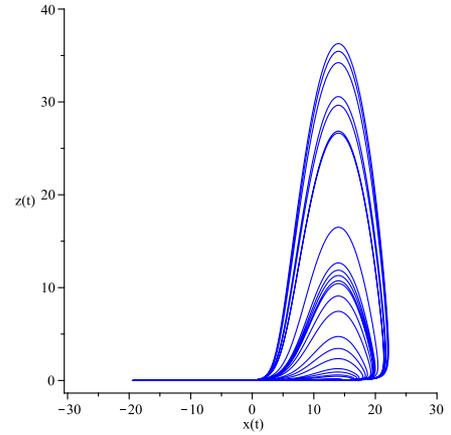}
	\label{fig:raxz}	
}
\\
\subfigure[]{
	\includegraphics[height=.24\textheight]{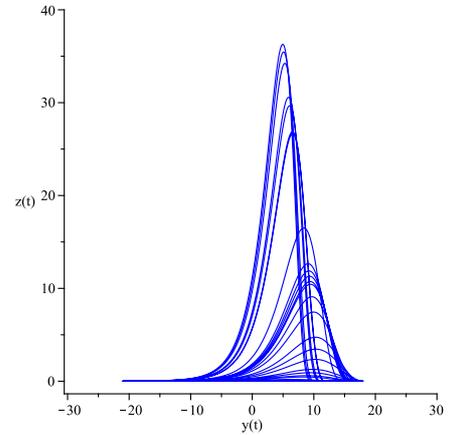}
	\label{fig:rayz}	
}

\end{center}	
	\caption{2-Dimensional plot of R{\"o}ssler Attractor: (a) x-y plane, (b) x-z plane, and (c) y-z plane}
	\label{fig:ra2d}
\end{figure}

\begin{figure}[ht]
\begin{center}
\subfigure[]{
	\includegraphics[height=.24\textheight]{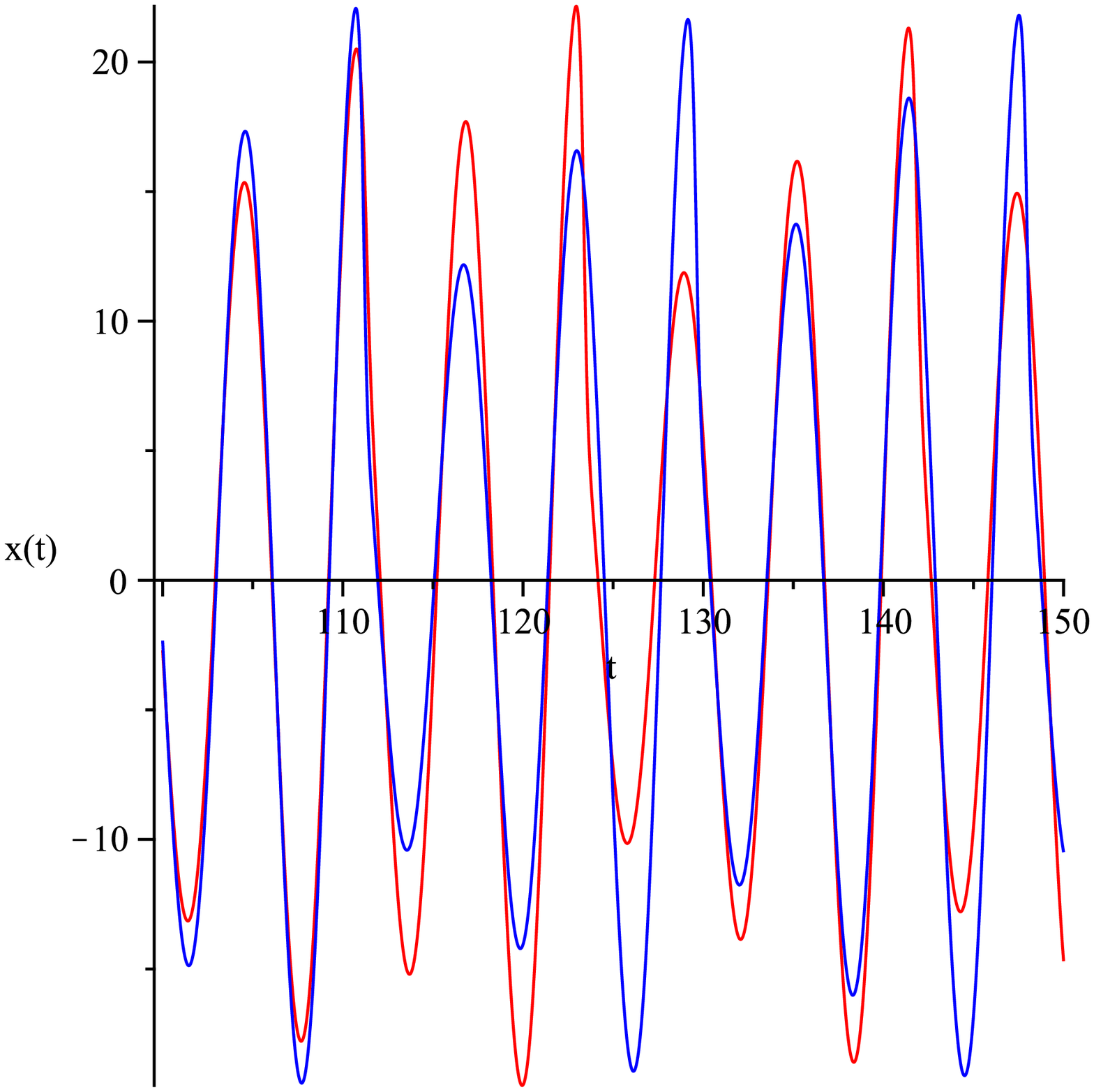}
	\label{fig:ratx}	
}
\\
\subfigure[]{
	\includegraphics[height=.25\textheight]{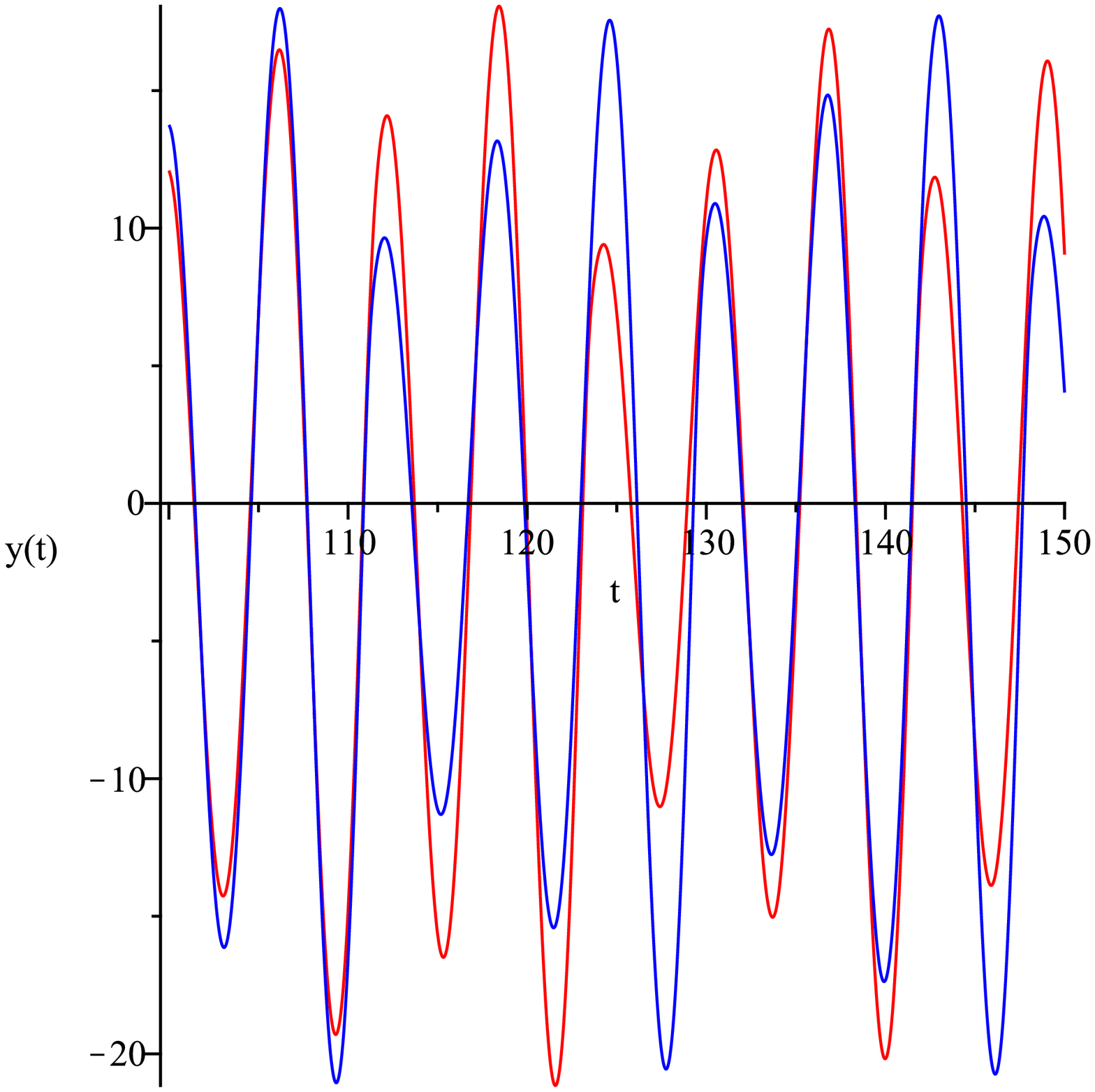}
	\label{fig:raty}	
}
\\
\subfigure[]{
	\includegraphics[height=.24\textheight]{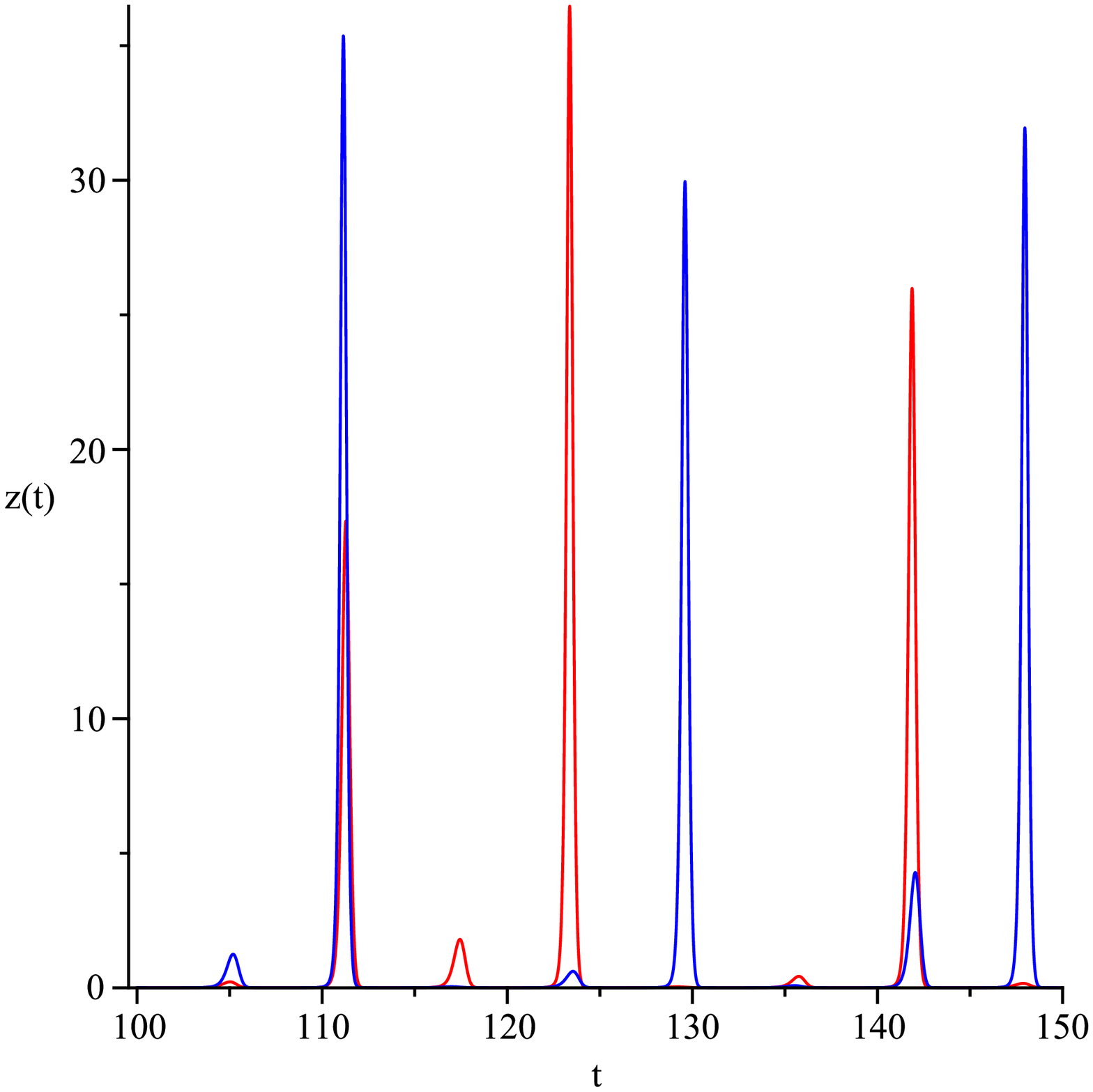}
	\label{fig:ratz}	
}

\end{center}	
	\caption{Time series plot of R{\"o}ssler Attractor: (a) t-x plane, (b) t-y plane, and (c) t-z plane. Shows sensitivity to initial conditions with $x_0=1$ (\textcolor{blue}{blue}) and  $x_0=1.001$ (\textcolor{red}{red}).}
	\label{fig:rats}
\end{figure}

The famous R\"osler attractor is presented in \cite{Rossler}, Eq. (\ref{eq:RosslerAttractor}): 
\begin{align}\label{eq:RosslerAttractor}
	\begin{aligned}
&\frac{dx}{dt}=-y-z \\
&\frac{dy}{dt}=x+ay \\
&\frac{dz}{dt}=b+z(x-c) \ ,
	\end{aligned}
\end{align} 
where the parameters $a$, $b$, and $c$ are positive real numbers. The system is chaotic when $a=0.2$, $b=0.2$, and $c=5.7$. Three-Dimensional model of R{\"o}ssler Attractor is illustrated in Fig. \ref{fig:ra3d}. Fig. \ref{fig:ra2d} shows 2-Dimensional plot of R{\"o}ssler Attractor. Sensitivity to initial conditions are shown if Fig. \ref{fig:rats}.

\subsection{Bent Boolean Functions}
In this subsection we refer to works of Cusick and St\u{a}nic\u{a} \cite{BS:TC}, Neumann \cite{Neumann}, Pommerening \cite{BS:P}, and Rothaus \cite{BS:R}.
\begin{defn}
A Boolean function $f$ in $n$ variables is map from $\mathbb{V}_n$ (the vector space on $n$ dimension) to the two-element Galois field $\mathbb{F}_2$. The (0,1)-sequence defined by $(f(\mathbf{x}_0),f(\mathbf{x}_1),\dots,f(\mathbf{x}_{2^{n-1}}))$ is called the truth table of $f$, where $\mathbf{v}_0=(0,\dots,0,0),\mathbf{v}_1=(0,\dots,0,1),\dots,\mathbf{v}_{2^{n-1}}=(1,\dots,1,1)$, ordered by lexicographical order.
\end{defn}

To each Boolean function $f:\mathbb{V}_n\rightarrow\mathbb{F}_n$ we associate \emph{sign} function, denoted by $\hat{f}:\mathbb{V}_n\rightarrow\mathbb{R}^*\subseteq\mathbb{C}^*$ and defined by $\hat{f}(\mathbf{x})=(-1)^{f(\mathbf{x})}$.

\begin{defn}
The Walsh transform of a function $f$ on $\mathbb{V}_n$ (with the values of $f$ taken to be real numbers 0 and 1) is the map $W(f):\mathbb{V}_n\rightarrow\mathbb{R}$, defined by 
\[
W(f)(\mathbf{w})=\sum_{\mathbf{x}\in\mathbb{V}_n}{f(\mathbf{x})(-1)^{\mathbf{w}\cdot\mathbf{x}}} \,
\]
which defines the coefficients of f with respect to the orthonormal basis of the group characters $Q_{\mathbf{x}}(\mathbf{w})=(-1)^{\mathbf{w}\cdot\mathbf{x}}$ (where $\mathbf{w}\cdot\mathbf{x}$ is the scalar product); $f$ can be recovered by the inverse Walsh transform
\[
f(\mathbf{x})=2^{-n}\sum_{\mathbf{w}\in\mathbb{V}_n}{W(f)(\mathbf{w})(-1)^{\mathbf{w}\cdot\mathbf{x}}} \,
\]
\end{defn}

\begin{defn}
A Boolean function $f$ in $n$ variables is called \emph{bent} if and only if the Walsh transform coefficients of $\hat{f}$ are all $\pm2^{n/2}$, that is, $W(\hat{f})^2$ is constant.
\end{defn}

Maiorana construction provides us with the following bent function 
\begin{equation}\label{eq:MaioranaNeumann}
x_0y_0\oplus ... \oplus x_{m-1}y_{m-1} \oplus x_0x_1 \cdots x_{m-1},
\end{equation}
where the Boolean function is presents by the polynomial $P(x_0,...,x_{m-1},y_0,...,y_{m-1})=R(x_0,...,x_{m-1})\oplus x_0y_0\oplus ... \oplus x_{m-1}y_{m-1}$ and its dual polynomial $P^{\star}(x_0,...,x_{m-1},y_0,...,y_{m-1})=R(x_0,...,x_{m-1})\oplus x_0y_0\oplus ... \oplus x_{m-1}y_{m-1}$, where $R=\mathbb{R}(m)$ is an arbitrary polynomials in $m$ variables.
\subsection{Proposed Bit Generator}
The novel algorithm is based on the following steps:
\begin{enumerate} []
{\setlength\itemindent{1cm}\item[\textbf{\textit{Step 1:}}] The initial values $x_0$, $y_0$, and $z_0$ from Eq. (\ref{eq:RosslerAttractor}) are determined.}

{\setlength\itemindent{1cm}\item[\textbf{\textit{Step 2:}}] With using the third-order Runge-Kutta method \cite{Farago2014}, we numerically compute the attractor from Eq. (\ref{eq:RosslerAttractor}) with step size $h=0.01$. It is iterated for $L_{1}$ times.}

{\setlength\itemindent{1cm}\item[\textbf{\textit{Step 3:}}] The iteration of the Eq. (\ref{eq:RosslerAttractor}) continues, and as a result, two real fractions $x_i$ and $y_{i}$, are generated and post-processed as follows:
\begin{equation*}\label{Post-proc}
	\begin{aligned}
s_{0} &= abs(integer(x_{i}\times 10^{7}))  \\
s_{1} &= abs(integer(y_{i}\times 10^{7})), \\
 	\end{aligned}
\end{equation*}  
where $integer(x)$ returns the integer part of $x$, truncating the value at the decimal point and $abs(x)$ returns the absolute value of $x$.}

{\setlength\itemindent{1cm}\item[\textbf{\textit{Step 4:}}] Generate a numeric vector $V=(s_{0}[0],...,s_{0}[p],s_{1}[0],...,s_{1}[p])$ that contains the digits of the  numbers $s_{0}$ and $s_{1}$.}

{\setlength\itemindent{1cm}\item[\textbf{\textit{Step 5:}}] Apply the vector $V$ to Maiorana function from Eq.  (\ref{eq:MaioranaNeumann})  to get a single output bit.}

{\setlength\itemindent{1cm}\item[\textbf{\textit{Step 6:}}] Return to Step 3 until the bit stream limit is reached.}
\end{enumerate}

The proposed bit generator is implemented in C++, using the following initial values: $x_0=0.1$, $y_0=0.15$, $z_0=0.01$, and $L_{1}=2000$.

\subsection{Key space evaluation}
The secret key space is composed by the four secret keys $x_0$, $y_0$, $z_0$, and $L_1$. With number of about $15$ decimal digits precision in IEEE double precision \cite{IEEE} the proposed key space is more than $2^{126}$, which is good enough against exhaustive key search \cite{AlvarezLi}.

\subsection{Statistical tests}
Three software test programs are used in order to measure the behaviour of the output binary streams.

The DIEHARD package \cite{DIE} includes 19 statistical tests: Birthday spacings, Overlapping 5-permutations, Binary rank (31 x 31), Binary rank (32 x 32), Binary rank (6 x 8), Bitstream, Overlapping-Pairs-Sparse-Occupancy, Overlapping-Quadruples-Sparse-Occupancy, DNA, Stream count-the-ones, Byte-count-the-ones, Parking lot, Minimum distance, 3D spheres, Squeeze, Overlapping sums, Runs (up and down), and Craps. The tests return $P-values$, which should be uniform in [0,1), if the input file contains pseudorandom numbers. The $P-values$ are obtained by $p=F(y)$, where $F$ is the assumed distribution of the sample random variable $y$, often the normal distribution.

The NIST software application \cite{BS:NIST} is a set of 15 statistical tests: Frequency (monobit), Block-frequency, Cumulative sums (forward and reverse), Runs, Longest run of ones, Rank, Fast Fourier Transform (spectral), Non-overlapping templates, Overlapping templates, Maurer’s "Universal Statistical", Approximate entropy, Random excursion, Random-excursion variant, Serial, and Linear complexity.

The testing process consists of the following steps:
\begin{enumerate} []
{\setlength\itemindent{1cm}\item[\textbf{\textit{Step 1:}}] State the null hypothesis. Assume that the zero/one sequence is random.}
{\setlength\itemindent{1cm}\item[\textbf{\textit{Step 2:}}] Compute a sequence test statistic. Testing is carried out at the bit level.}
{\setlength\itemindent{1cm}\item[\textbf{\textit{Step 3:}}] Compute the $P-value$, $P-value \in [0,1]$.}
{\setlength\itemindent{1cm}\item[\textbf{\textit{Step 4:}}] Fix $\alpha$, where $\alpha \in [0.0001,0.01]$. Compare the $P-value$ to $\alpha$. $Success$ is declared whenever $P-value \geq \alpha$; otherwise, $failure$ is declared.}
\end{enumerate}
The NIST package calculates the proportion of sequences that pass the particular tests. The range of acceptable proportion is determined using the confidence interval defined as,
\[
\hat{p}\pm 3\sqrt{\frac{\hat{p}(1-\hat{p})}{m}},
\]
where $\hat{p}=1-\alpha$, and $m$ is the number of binary tested sequences. NIST recommends that, for these tests, the user should have at least 1000 sequences of 1000000 bits each. In our setup $m=1000$. Thus the confidence interval is
\[
0.99\pm 3\sqrt{\frac{0.99(0.01)}{1000}}=0.99 \pm 0.0094392.
\]
The proportion should lie above 0.9805607 with exception of Random excursion and Random excursion variant tests. These two tests only apply whenever the number of cycles in a sequence exceeds 500. Thus the sample size and minimum pass rate are dynamically reduced taking into account the tested sequences.

The distribution of $P-values$ is examined to ensure uniformity. The interval between 0 and 1 is divided into 10 subintervals. The $P-values$ that lie within each subinterval are counted. Uniformity may also be specified through an application of a $\chi^2$ test and the determination of a $P-value$ corresponding to the goodness-of-fit distributional test on the $P-values$ obtained for an arbitrary statistical test, $P-value$ of the $P-values$. This is implemented by calculating
\[
\chi^2=\sum_{i=1}^{10}{\frac{(F_i-s/10)^2}{s/10}},
\]
where $F_i$ is the number of $P-values$ in subinterval $i$ and $s$ is the sample size. A $P-value$ is computed such that $P-value_T=IGAMC(9/2,\chi^2/2)$, where $IGAMC$ is the complemented incomplete gamma statistical function. If $P-value_T \geq 0.0001$, then the sequences can be considered to be uniformly distributed.

The ENT package \cite{ENT} performs 6 tests to sequences. They are Entropy, Optimum compression, ${\chi}^2$ distribution, Arithmetic Mean value, Monte Carlo Value for $\pi$, and Serial Correlation Coefficient. The sequences of bytes are stored in files. The suite outputs the results of those tests. We tested output stream of 125000000 bytes of the novel pseudorandom number generator.

The test results are given in Table \ref{tab:tabDIEHARD}, Table \ref{tab:tabNIST}, and Table \ref{tab:tabENT}, respectively.
All of statistical tests are passed successfully.
\begin{table}[!htbp]
	\caption{DIEHARD statistical test results for two 80 million bits sequences generated by the proposed generator}
	\label{tab:tabDIEHARD}
	\footnotesize
	\centering
	\begin{tabular}{|l|c|}
		\hline
		\tablehead{1}{\textbf{DIEHARD}} 
		& \tablehead{1}{\textbf{Proposed Generator}}\\ 
		\tablehead{1}{\textbf{statistical test}}
		& \tablehead{1}{\textit{\textbf{P-value}}}\\ 
		\hline
Birthday spacings         & 0.593701  \\
Overlapping 5-permutation & 0.395409  \\
Binary rank (31 x 31)     & 0.746323  \\
Binary rank (32 x 32)     & 0.955445  \\
Binary rank (6 x 8)       & 0.470244  \\
Bitstream                 & 0.479493  \\
OPSO                      & 0.522760  \\
OQSO                      & 0.569775  \\
DNA                       & 0.509277  \\
Stream count-the-ones     & 0.213018  \\
Byte count-the-ones       & 0.612866  \\
Parking lot               & 0.428480  \\
Minimum distance          & 0.479449  \\
3D spheres                & 0.470514  \\
Squeeze                   & 0.935011  \\
Overlapping sums          & 0.616515  \\
Runs up                   & 0.317489  \\
Runs down                 & 0.347915  \\
Craps                     & 0.587091  \\
		\hline
	\end{tabular}

\end{table}

\begin{table}[!htbp]
\caption{NIST Statistical test suite results for 1000 sequences of size $10^6$-bit each generated by the proposed generator}
\label{tab:tabNIST}
\footnotesize
\centering
\begin{tabular}{|l|r|r|}
\hline
	\tablehead{1}{\textbf{NIST}}
  & \multicolumn{2}{c|}{\textbf{Proposed Generator}}\\ 
\cline{2-3}
	\tablehead{1}{\textbf{statistical test}}
  & \tablehead{1}{\textit{\textbf{P-value}}}
  & \tablehead{1}{\textbf{Pass rate}}\\ 
\hline
Frequency (monobit) 		& 0.896345	& 991/1000 \\
Block-frequency				& 0.348869	& 988/1000 \\
Cumulative sums (Forward)	& 0.095426  & 992/1000 \\
Cumulative sums (Reverse)	& 0.632955  & 993/1000 \\
Runs						& 0.195864  & 992/1000 \\
Longest run of Ones			& 0.597620	& 990/1000 \\
Rank						& 0.587274  & 992/1000 \\
FFT							& 0.849708  & 987/1000 \\
Non-overlapping templates	& 0.505628  & 990/1000 \\
Overlapping templates		& 0.308561  & 986/1000 \\
Universal					& 0.474986  & 988/1000 \\
Approximate entropy			& 0.973718  & 987/1000 \\
Random-excursions			& 0.476145  & 629/635 \\
Random-excursions Variant	& 0.502142	& 630/635 \\
Serial 1					& 0.707513  & 997/1000 \\
Serial 2					& 0.729870  & 987/1000 \\
Linear complexity			& 0.096578  & 993/1000 \\\hline
\end{tabular}
\end{table}

\begin{table}[!htbp]
\caption{ENT statistical test results for two 80 million bits sequences generated by the proposed generator.}
\label{tab:tabENT}
\footnotesize
\centering
\begin{tabular}{|l|l|}
\hline
	\tablehead{1}{\textbf{ENT}}
  & \tablehead{1}{\textbf{Proposed Generator}}\\ 
	\tablehead{1}{\textbf{statistical test}}
  & \tablehead{1}{\textbf{results}}\\ 
\hline
Entropy                   & 7.999998 bits per byte       \\
Optimum compression       & OC would reduce the size of  \\
					      & this 125000000 byte file     \\
					      & by 0 $\%$.                   \\
${\chi}^2$ distribution   & For 125000000 samples is     \\
					      & 271.65, and randomly would   \\
					      & exceed this value 22.62 $\%$ \\					      
					      & of the time.                 \\
Arithmetic mean value     & 127.4991  (127.5 = random)   \\
Monte Carlo $\pi$ estim.  & 3.141880562 (error 0.01 $\%$)\\
Serial correl. coeff.     & -0.000089                    \\
					      & (totally uncorrelated = 0.0) \\
\hline
\end{tabular}
\end{table}

\section{Conclusions and Future Work}

A novel pseudorandom number generator based on a chaotic map is proposed in this article. The proposed algorithm combines R{\"o}ssler attractor, and Maiorana bent Boolean function. 

An accurate security analysis on the novel scheme is given. 
Based on the results, we can conclude that the proposed pseudorandom number generation algorithm is acceptable for the secure data encryption. 

We intend to use this algorithm for Galois field based encryption \cite{TashevaMSE}.


%

\section*{Acknowledgment}

The authors are grateful to the anonymous referees for valuable and helpful comments.

The authors would like to thank Borislav Bedzhev and Stanimir Stanev for their comments and suggestion on earlier drafts of this paper.

This work is partially supported by the Scientific research fund of Konstantin Preslavski University of Shumen under the grant No. RD-08-121\slash 06.02.2017.

\ifCLASSOPTIONcaptionsoff
  \newpage
\fi

\end{document}